\documentclass{aipproc}

\layoutstyle{6x9}

\usepackage{graphicx}
\begin{document}
\title{Is HBT really puzzling?}


\author{Scott Pratt and Daniel Schindel}{
address={Department of Physics and Astronomy,
Michigan State University\\ East Lansing, Michigan 48824}
email={pratts@pa.msu.edu}
}

\begin{abstract}
Two-particle correlations from RHIC have provided a surprising snapshot of the final state at RHIC. In this talk I discuss the nature of the HBT puzzle and attempt to delineate several factors which might ultimately resolve the issue.
\end{abstract}

\date{\today}

\maketitle

Correlation, or HBT, measurements provide our best insight into the space-time development of relativistic heavy-ion collisions \cite{Lisa:2005dd}. Immediately after the first two-pion correlations from RHIC were presented in 2001, the term "HBT puzzle" came into common usage. The term was inspired by the failure of some of our most sophisticated dynamical models of the collision. However, there is no problem in describing the data with simple parameterizations of the break-up space-time profile. In this talk, I will review one of these simple parameterizations, the blast-wave model, to illustrate the unsettling aspects of the inferred parameters. I then discuss several factors which might ultimately lead to a more satisfying interpretation of the data.

Numerous parameterizations of the breakup space-time geometry can be applied to fit correlation data from RHIC. Among these are the Buda-Lund model \cite{Csanad:2003qa} and numerous variations of the blast-wave model \cite{Retiere:2003kf,Schnedermann:1993ws,Kisiel:2005hn}. Minimal blast wave models incorporate four common parameters: the breakup temperature $T$, the radius of the fireball surface $R$, the collective velocity at the surface $v_\perp$, and the breakup time $\tau$, which is related to the collective-velocity gradient along the beam direction by the relation $v_z=z/t$ by virtue of assuming boost invariance. Additionally, chemical potentials can be added to normalize the spectra. Parameters for the surface diffuseness or the temporal duration of the emission $\Delta\tau$ are often included. The Buda-Lund model also allows for a temperature gradient.

\begin{figure}
\centerline{\includegraphics[width=0.6\textwidth]{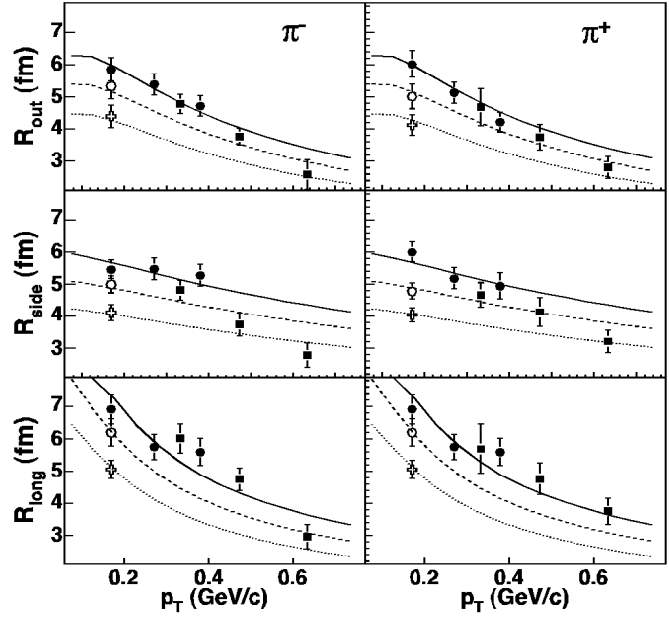}}
\caption{\label{fig:blast}
Blast wave fits to source sizes from Lisa and Retiere \cite{Retiere:2003kf}.
}
\end{figure}

Pionic observables do not tightly constrain the parameters by themselves \cite{Tomasik:2003gt}, but after fitting the spectra of heavier particles the following parameters emerge from blast-wave fits: $T\sim 110$ MeV, $v_\perp\sim 0.7c$, $R\sim 12-13$ fm, and $\tau\sim 9$ fm/$c$. The duration of the emission appears to be sudden, less than 5 fm/$c$, and roughly consistent with zero. The quality of the fits with data are illustrated in Fig. \ref{fig:blast}.

On inspection, there are two puzzling aspects of these numbers. First, the breakup density is remarkably high. If one takes the 1000 hadrons per unity rapidity observed at RHIC and divide it by the volume $V=\tau \pi R^2$, one obtains densities higher than nuclear matter density, in the range of 0.18 fm$^{-3}$. Assuming cross sections of 20 mb, estimates of the mean free path are in the neighborhood of 2.5 fm, which is much smaller than the 25 fm diameter of the fireball. The second surprising aspect of the fit is that it implies the fireball, whose original diameter was 6 fm, expanded by 7 fm in 10 fm/$c$. Given the surface velocity was 0.7$c$, this implies the surface unphysically accelerates instantaneously to its final velocity.

The density, though surprising, can be readily explained. First, due to collective expansion, the density will have fallen substantially by the time particles move a few fm. Secondly, at the time of breakup, the expansion has become isotropic. For non-relativistic Gaussian-like expansions of particles of the same mass, collisions do not alter the outgoing phase space density, and thus cease to have an effect on the overall source size \cite{Chojnacki:2004ec}. Thus, the phase space distribution tends to effectively freeze out before particles have their last collisions. In fact, microscopic simulations incorporating cross sections of tens of mb often yield source sizes smaller than those measured at RHIC. The freezing-out of the source size can be thought of as a competition between the cooling, which pushes the system toward smaller source sizes and the growth of the overall source. For the qualifiers mentioned above, these two balance perfectly. In practice, collisions of pions with heavier particles tend to reduce the pion source and increase the apparent source size of heavier particles such as protons, as entropy moves from pions to protons. In fact, source sizes from microscopic models usually vary remarkably little as a function of the assumed cross section \cite{Pratt:1998gt}.

\begin{figure}
\centerline{\includegraphics[width=0.5\textwidth]{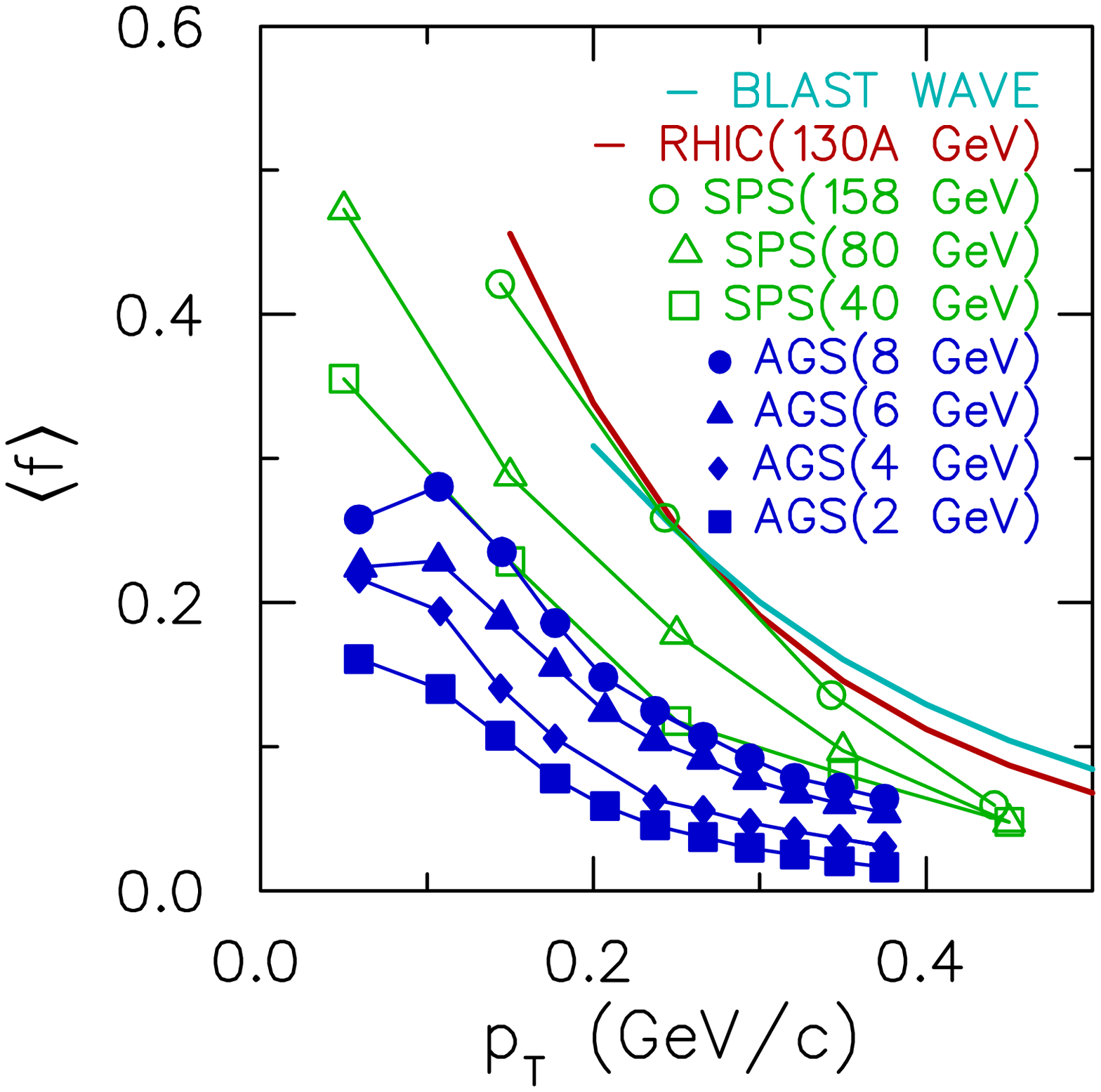}
\hspace*{0.05\textwidth}\includegraphics[width=0.45\textwidth]{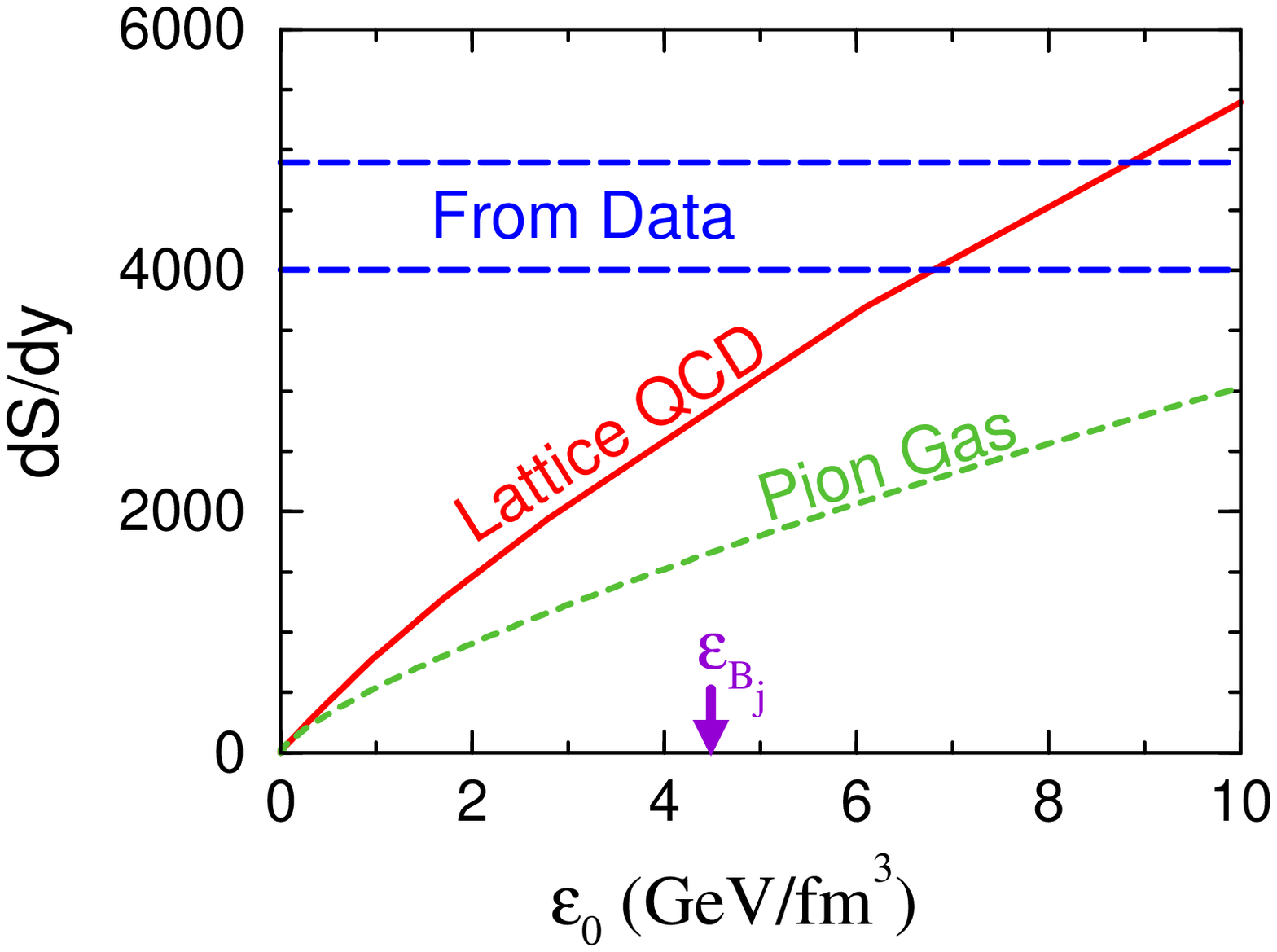}}
\caption{\label{fig:psdentropy}
Average pionic phase space densities as a function of $p_t$ (left panel) rise with increasing beam energy, but saturate at upper SPS energies. As a function of the initial energy density at 1 fm/$c$, expected entropies per unit rapidity can be calculated given the equation of state (right panel). A blast-wave model with a chemical potential of 70 MeV for pions roughly reproduces the RHIC results. The Bjorken estimate probably underestimates the true energy density by $\sim 50\%$, thus the lattice equation of state seems consistent with the experimentally extracted $dS/dy$ which is slightly above 4000.
}
\end{figure}

Source sizes are constrained by entropy, which can be calculated from final-state phase space densities, which can in turn be extracted from experiment given measured radii,
\begin{eqnarray}
\bar{f}({\bf p})&\equiv&
\frac{\int d^3r [f({\bf p},{\bf r})]^2}{\int d^3r f({\bf p},{\bf r})}\\
\nonumber
&=&\frac{\pi^{3/2}}{(2S+1)}\cdot \frac{dN/d^3p}
{R_{\rm out}R_{\rm long}R_{\rm side}}\\
\frac{dS}{dy}&\approx&2\pi\int p_tdp_t~E\frac{E}{d^3p}
\left[\frac{5}{2}-\frac{3}{2}\log(2)-\log(\bar{f}(p_t))\pm\frac{1}{2^{3/2}}\bar{f}(p_t)\right]
\end{eqnarray}
These expressions were applied in \cite{Pal:2003rz} to extract entropy and phase space density in central 130$A$ GeV collisions at RHIC. Phase space densities at this and lower energies from \cite{Lisa:2005dd} are displayed in Fig. \ref{fig:psdentropy}. Note that the phase space densities tend to rise, then saturate near the top SPS energies. At this point the phase space density, which is approximately frozen after chemical freezeout in an isentropic expasion, saturates when the entirety of the fireball surpasses temperatures of 170 MeV \cite{Akkelin:2005ms}, the temperature for which particle ratios suggest chemical compositions have frozen out \cite{Braun-Munzinger:2001ip}.

Pionic source volumes are a factor of two smaller than those predicted by hydrodynamic models based on lattice-gauge-theory-inspired equations of state. This represents a deficit of $\ln(2)$ units of entropy per pion, which might suggest that the equations of state used for the hydrodynamic calculations in \cite{Soff:2000eh} is excluded. However, an analysis of the entropy carried by all the particles is consistent with entropy expected from lattice calculations \cite{Pal:2003rz} as shown in Fig. \ref{fig:psdentropy}. Since baryons tend to carry twice as much entropy per particle as pions, a modest increase in the population of baryons is able to absorb the $\sim 20\%$ shortfall in pionic entropy.  If one knows $s(\epsilon)$, the entropy density as a function of the energy density, one can also derive the pressure $P(\epsilon)$ from thermodynamic identities. A higher-entropy equation of state, i.e., one where the effective number of degrees of freedom is higher, must also have lower pressure than an equation of state whose entropy rises more slowly with energy. Since entropy can increase during the expansion, the fact that the measured entropy is consistent with the lattice equation of state places a lower-bound on the stiffness of the equation of state. A very stiff equation of state, like the pion gas example illustrated in Fig. \ref{fig:psdentropy}, could in principle be consistent with the final-state entropy, but would require the entropy to double during the expansion. Most mechanisms for generating entropy such as viscosity and shocks are expected to provide $\sim 10\%$ increases in entropy.

\begin{figure}
\centerline{\includegraphics[width=0.5\textwidth]{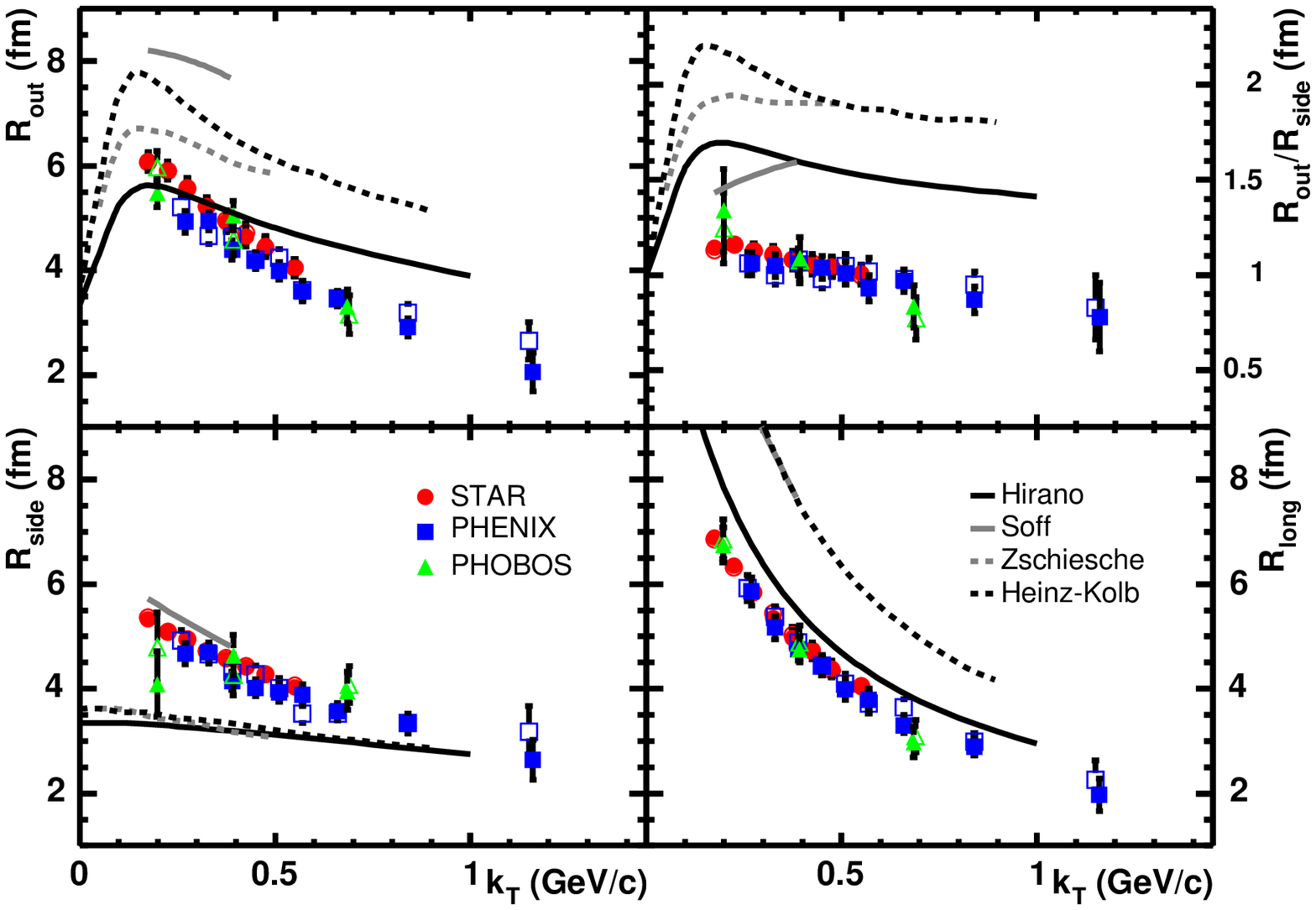}
\includegraphics[width=0.5\textwidth]{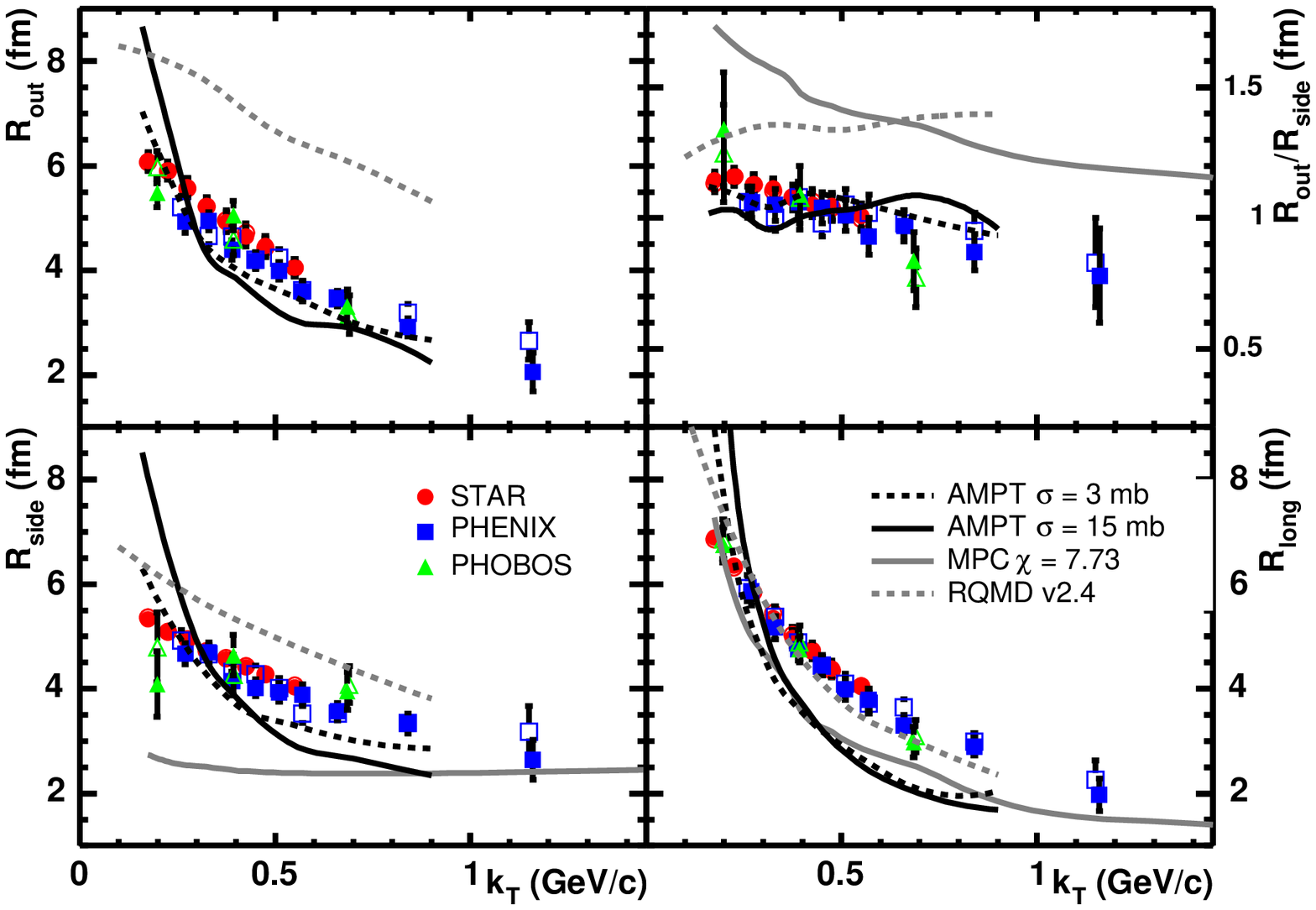}}
\caption{\label{fig:models}
Dimensions of several models from identical-pion correlations compared to RHIC data as a function of the average pair momentum $k_t$. A basic trend is apparent that those models with soft equations of state overestimate the source sizes while the stiffest example, a pion gas (MPC) grossly underestimates the source sizes. None of the models quantitatively describe the $k_t$ dependence of all the dimensions.}
\end{figure}

Unfortunately, there are no systematic studies of the dependence of source radii as a function of the equation of state. However, a compilation of several hydrodynamic and microscopic transport models presented in Fig. \ref{fig:models} shows a trend of stiffer equations of state resulting in smaller radii, as expected. By incorporating more resonances, or by including strings, microscopic models effectively lower the pressure. The largest source sizes resulted from hydrodynamic calculations using an equation of state with a large latent heat. Although it is premature, one might conclude that the equation of state appears somewhat stiffer than some of the lattice models, but not nearly so stiff as that of a pion gas. Taken as a whole the model calculations are encouraging as they illustrate the sensitivity of correlation meausurements to the equation of state.

Even though the average source sizes can be reproduced with some of the transport models represented in Fig. \ref{fig:models}, none of the models does a good job of reproducing the $p_t$ dependence for all three dimensions. Part of this difficulty lies in the fact that, as stated earlier, models tend to require some time to accelerate transversely and thus have difficulty reproducing the final-state geometry described by the blast-wave model. This aspect of the HBT puzzle might be resolved by some combination of the following improvements or alterations of dynamic models:
\begin{enumerate}
\item Reduced emissivity. Sequential evaporation from the surface tends to increase $R_{\rm out}/R_{\rm side}$ beyond the value of unity observed. Since pions with $p_t>100$ MeV/$c$ move faster than the surface, those pions emitted earlier get ahead of other pions and extend the shape of the phase space distribution. Any mechanism to reduce the emissivity of the surface, such as super-cooling \cite{Csorgo:1994dd}, would help the system reach the sudden freeze-out picture of a blast-wave model.
\item Non-infinite longitudinal extent. Many of the hydrodynamic models assume boost invariance. A finite extent along the beam direction lowers $R_{\rm long}$, and since $R_{\rm long}$ is related to the inferred lifetime, accounting for the finiteness should lead to longer lifetimes which gives the matter more time to expand transversely to the large 13 fm size.
\item Longitudinal acceleration. Boost invariant models also neglect acceleration along the beam axis. Accounting for this acceleration alters the connection between the velocity gradient and the lifetime, $dv_z/dz\sim 1/\tau$, and should result in somewhat longer times \cite{renk_thisproc}.
\item Shear viscosity. The high velocity gradient along the beam axis at early times can lead to shear effects which lower the pressure along the beam axis  while raising the transverse pressure \cite{Teaney:2004qa,Muronga:2004sf}. This effect could be especially significant if the early stage is well described by longitudinal classical fields. For non-interacting electric fields along the $z$ axis, the components of the stress energy tensor are $T_{xx}=T_{yy}=\epsilon$, $T_{zz}=-\epsilon$.
\item Refraction through the mean field. An attractive mean field for pions could refract trajectories and lead to phase space distributions whose $R_{\rm out}$ dimension exceeds the physical size of the emitting source \cite{Cramer:2004ih,Miller:2005ji,Pratt:2005hn}, especially the $R_{\rm side}$ projection.
\end{enumerate}
None of these effects are expected to represent more than a 10\% correction, except for the refractive effects at very low $p_t$, but they all push the analysis of the data toward a more physical interpretation. Thus, the HBT puzzle might well have originated from a conspiracy of effects, each of which borders on being negligible when considered by itself.

Unraveling the space-time picture of these collisions will undoubtedly benefit from improved measurement, especially the measurement of other correlations besides pion correlations \cite{Pratt:2003ar,danielewicz_thisproc}. However, the main impediment toward resolving correlation measurements remains the lack of a cohesive well-tested modeling framework. For instance, there is the need of a detailed comparison of hydrodynamic and Boltzmann approaches using the same effective equations of state, so that the role of viscosity, and the sensitivity to breakup conditions might be quantified. The random comparison of different models, each run with different initial conditions, unknown equations of state and viscosities, untested breakup criteria, and varying chemistries, precludes robust or rigorous scientific conclusions being generated from comparison with data. But, despite the daunting modeling challenge that lies ahead, the analyses performed thus far have demonstrated a keen sensitivity between the bulk properties used in the calculations and experimental correlation analyses.

\begin{theacknowledgments}
Support was provided by the U.S. Department of Energy, Grant No. DE-FG02-03ER41259.
\end{theacknowledgments}

\end{document}